\documentclass[
11pt,%
tightenlines,%
twoside,%
onecolumn,%
nofloats,%
nobibnotes,%
nofootinbib,%
superscriptaddress,%
noshowpacs,%
centertags]%
{revtex4}

\usepackage{ljm}
\usepackage{graphicx}
\usepackage{algorithm}
\usepackage{algorithmic}
\usepackage{cleveref}
\usepackage{tabularx}
\usepackage{booktabs}
\usepackage{subfigure}
\usepackage{listings}
\usepackage{fancyvrb}
\usepackage{xspace}

\newcolumntype{Y}{>{\raggedleft\arraybackslash}X}

\newcommand{\CXX}{{C\nolinebreak[4]\hspace{-.05em}\raisebox{.3ex}{\tiny\bf ++}}\xspace}
\newcommand{\code}[1]{\lstinline|#1|}

\crefname{lstlisting}{listing}{listings}
\Crefname{lstlisting}{Listing}{Listings}

\begin{document}

\title{Efficient solution of 3D elasticity problems with smoothed aggregation
algebraic multigrid and block arithmetics}

\author{\firstname{D.~E.}~\surname{Demidov}}
\email[E-mail: ]{dennis.demidov@gmail.com}

\affiliation{
    Kazan Branch of Joint Supercomputer Center,
    Scientific Research Institute of System Analysis, \\
    the Russian Academy of Sciences;
    2/31, Lobachevskii str., Kazan 420111 Russia}

\firstcollaboration{(Submitted by A. M. Elizarov)}

\received{February 17, 2022}

\begin{abstract}
    Efficient solution of 3D elasticity problems is an important part of many
    industrial and scientific applications. Smoothed aggregation algebraic
    multigrid using rigid body modes for the tentative prolongation operator
    construction is an efficient and robust choice for the solution of linear
    systems arising from discretization of elasticity equations. The system
    matrices on every level of the multigrid hierarchy have block structure, so
    using block representation and block arithmetics should significantly
    improve the solver efficiency. However, the tentative prolongation operator
    construction may only be done using scalar representation. The paper
    proposes a couple of practical approaches for enabling the use of block
    arithmetics with smoothed aggregation algebraic multigrid based on the
    open-source AMGCL library. It is shown on the example of two real-world
    model problems that the suggested improvements may speed up the solution by
    50\% and reduce the memory requirements for the preconditioner by 30\%. The
    implementation is straightforward and only requires a minimal amount of
    code.
\end{abstract}

\subclass{35-04, 65-04, 65Y05, 65Y10, 65Y15, 97N80}
\keywords{Smoothed aggregation AMG, 3D elasticity, rigid body modes,
nullspace components, block arithmetics.}

\maketitle

\section{Introduction}

Solution of large sparse linear systems obtained by discretization of 3D
elasticity equations is an important part in many industrial and scientific
simulation applications.  Iterative solvers from Krylov subspace preconditioned
with smoothed aggregation algebraic multigrid (AMG) are considered to be an
efficient and robust choice for solving such systems~\cite{brandt1985algebraic,
Trottenberg2001, Stuben1999}.  The system matrices resulting from the 3D
elasticity equations usually have block structure with small $3 \times 3$
blocks. It has been shown in~\cite{DEMIDOV2021101285} that using block
representation and block arithmetics for such systems may reduce both the
computation time and the memory required for the preconditioner. In order to
achieve good convergence, the AMG preconditioner or, more specifically, the
tentative prolongation operator has to be constructing using geometry-specific
information in form of zero-energy modes, or near nullspace components, which
in case of elasticity problems consist of the six rigid body
modes~\cite{VanekBrezinaMandel2001}. However, the tentative prolongation
operator construction is expressed in terms of scalar arithmetics, which makes
it hard to use block representation of the system matrix and block arithmetics
during the preconditioner setup in general.

This paper proposes a couple of practical approaches allowing to combine the
use of zero-energy modes with the block arithmetics in the smoothed aggregation
AMG setup. The first approach converts the constructed matrices to the block
representation as the final step of the setup.  This allows to speed up the
solution phase of the algorithm, but does not improve the efficiency of the
method initialization. The second approach constructs the transfer operators
using the near nullspace components in the scalar space, while keeping the rest
of the setup in the block domain. On the example of two real-world elasticity
problems it is shown that the latter approach allows to speed up the overall
computation by 40--50\%, and reduce the memory footprint of the preconditioner
by about 30\%. The suggested modifications are implemented as part of the
open-source \CXX library AMGCL~\cite{Demidov2019, Demidov2020}.

The rest of the paper is organized as follows: \Cref{sec:saamg} outlines the
smoothed aggregation AMG and describes the block representation of the
matrices; \Cref{sec:nsblock} shows how block arithmetics may be used in the
smoothed aggregation AMG setup when zero-energy modes are provided by the user;
\Cref{sec:experiments} provides the benchmarking results for the proposed
approaches on the example of two real-world 3D elasticity models.

\section{Smoothed aggregation algebraic multigrid} \label{sec:saamg}

Algebraic multigrid method~\cite{brandt1985algebraic, Stuben1999} solves a
system of linear algebraic equations \begin{equation} \label{eq:auf} Au = f,
\end{equation} where $A$ is a square matrix. Multigrid methods are based on
recursive application of the two-grid scheme, which combines \emph{relaxation}
and \emph{coarse grid correction}. Relaxation, or smoothing iteration $S$, is a
simple iterative method, such as damped Jacobi or Gauss--Seidel
iteration~\cite{barrett1994templates}. Coarse grid correction solves the
residual equation on the coarser grid, and improves the fine-grid approximation
with the interpolated coarse-grid solution. Transfer between the grids is
described with the \emph{transfer operators} $P$ (\emph{prolongation}) and $R$
(\emph{restriction}).

In geometric multigrid methods the grid hierarchy, the matrices $A_i$ and
operators $P_i$ and $R_i$ on each level of the hierarchy are supplied by the
user based on the problem geometry. In algebraic multigrid methods the grid
hierarchy and the transfer operators are in general constructed automatically,
based only on the algebraic properties of the matrix~$A$.

When solving elasticity problems, it is important to provide the rigid body
modes, or the zero-energy modes, or in more generic terms, the near nullspace
components for the linear system. The components are used during construction
of the tentative prolongation operator in order to ensure optimal convergence
of the solver. The near nullspace components are problem-specific and should be
supplied by the user. In case of 3D elasticity, the six rigid body modes are
computed from the discretization grid coordinates and provided in form of a $n
\times 6$ matrix $B$~\cite{VanekBrezinaMandel2001}.

\begin{algorithm}
    \caption{Smoothed aggregation AMG setup.}%
    \label{alg:setup}
    \begin{algorithmic}
        \STATE{Start with the square system matrix $A_1 = A$ and the
        matrix $B_1 = B$ containing the near nullspace components.}
        \WHILE{the matrix $A_i$ is too large to be solved directly}
            \STATE{
                Construct the prolongation operator $P_i$ and
                the next level matrix $B_{i+1}$ from $A_i$ and $B_i$,\\
                \quad obtain restriction operator $R_i = P_i^T$.}
            \STATE{Construct the smoother $S_i$ from $A_i$.}
            \STATE{
                Construct the coarser system using Galerkin operator:
                $A_{i+1} = R_i A_i P_i$.}
        \ENDWHILE
        \STATE{Construct a direct solver for the coarsest system $A_L$.}
    \end{algorithmic}
\end{algorithm}

\Cref{alg:setup} describes the setup phase of the smoothed aggregation AMG
method. Here, on each level of the AMG hierarchy, the prolongation operator
$P_i$ is constructed from the system matrix $A_i$ and the matrix $B_i$
containing the near nullspace components for the system. A common choice for
the restriction operator $R_i$ is the transpose of the prolongation operator
$R_i=P_i^T$. The next coarser level of the AMG hierarchy is fully defined by
the transfer operators $P_i$ and $R_i$.  After the AMG hierarchy has been
constructed, it is used as preconditioner with a Krylov subspace iterative
solver, where a single V-cycle shown in \Cref{alg:vcycle} is used as the
preconditioning step.

\begin{algorithm}
    \caption{AMG V-cycle.}
    \label{alg:vcycle}
    \begin{algorithmic}
        \STATE{Start at the finest level with an initial approximation
        $u_1 = u^0$.}
        \WHILE{not converged}
            \FOR{each level of the hierarchy, finest-to-coarsest}
                \STATE{Apply a couple of smoothing iterations to the current
                       solution: $u_i = S_i(f_i, u_i)$.}
                \STATE{Find residual $e_i = f_i - A_i u_i$ and
                       restrict it to the RHS on the coarser level:
                       $f_{i+1} = R_i e_i$.}
            \ENDFOR
            \STATE{Solve the coarsest system directly:
                   $u_L = A_L^{-1} f_L$.}
            \FOR{each level of the hierarchy, coarsest-to-finest}
                \STATE{Update the current solution with the interpolated
                       solution from the coarser level:
                       $u_i = u_i + P_i u_{i+1}$.}
                \STATE{Apply a couple of smoothing iterations to the current
                       solution: $u_i = S_i(f_i, u_i)$.}
            \ENDFOR
        \ENDWHILE
    \end{algorithmic}
\end{algorithm}

The AMGCL opensource \CXX library (published at
https://github.com/ddemidov/amgcl under permissive MIT license) provides a
flexible and extensible AMG implementation~\cite{Demidov2019,Demidov2020}. It
has a minimal set of dependencies, targets both shared and distributed memory
machines, and supports modern many-core architectures.  The value type concept
of the AMGCL library allows to generalize the provided algorithms for complex
or non-scalar systems.  A value type defines several overloads for common math
operations and is specified as a template parameter for a backend, where
backend is a class that defines matrix and vector types and implements parallel
primitives that are used during the solution phase of the algorithm. Most
often, a value type is simply a plain \code{double} or \code{float} scalar, but
it is also possible to use small statically sized matrices when the system
matrix has block structure, which decreases the setup time and memory footprint
of the algorithm, increases cache locality and may improve convergence rate in
some cases~\cite{gupta2010adaptive,DEMIDOV2021101285}.

Switching to a block-valued backend has several advantages. First, the block
representation of the matrix is more efficient memory-wise. This is
demonstrated on the example of the following sparse matrix that has a $2
\times 2$ block structure:
\begin{equation*}
    \begin{bmatrix}
        0.71 & 0.65 & 0.26 & 0.79 &      &      \\
        0.54 & 0.37 & 0.17 & 0.62 &      &      \\
             &      & 0.89 & 0.05 &      &      \\
             &      & 0.27 & 0.15 &      &      \\
             &      &      &      & 0.52 & 0.34 \\
             &      &      &      & 0.45 & 0.64
    \end{bmatrix}
\end{equation*}
Below is the standard scalar representation of the matrix in the CSR format:

\begin{Verbatim}[samepage=true]
ptr=[0, 4, 8, 10, 12, 14, 16]
col=[0, 1, 2, 3, 0, 1, 2, 3, 2, 3, 2, 3, 4, 5, 4, 5]
val=[0.71, 0.65, 0.26, 0.79, 0.54, 0.37, 0.17, 0.62,
     0.89, 0.05, 0.27, 0.15, 0.52, 0.34, 0.45, 0.64]
\end{Verbatim}
Compare it to the block-valued representation of the same matrix:
\begin{Verbatim}[samepage=true]
ptr=[0, 2, 3, 4]
col=[0, 1, 1, 2]
val=[[0.71, 0.65; 0.54, 0.37], [0.26, 0.79; 0.17, 0.62],
     [0.89, 0.05; 0.27, 0.15], [0.52, 0.34; 0.45, 0.64]]
\end{Verbatim}
With the block representation, the matrix has twice fewer rows and columns, and
four times fewer logical non-zero values.  This means that the matrix
representation needs twice less memory to store the \code{ptr} array, and four
times less memory for the \code{col} array. Since most of the parallel
primitives used in the AMG V-cycle and in iterative solvers in general are
memory-bound, this improves the performance of the solution phase. The reduced
logical size of the matrix also simplifies and speeds up the algorithm setup.
Finally, using the block arithmetics improves cache efficiency, since block
elements may be loaded and used all at once.

\begin{figure}
    \begin{center}
        \subfigure[The system matrix at the finest level of the AMG hierarchy.]{%
            \includegraphics[width=0.48\linewidth]{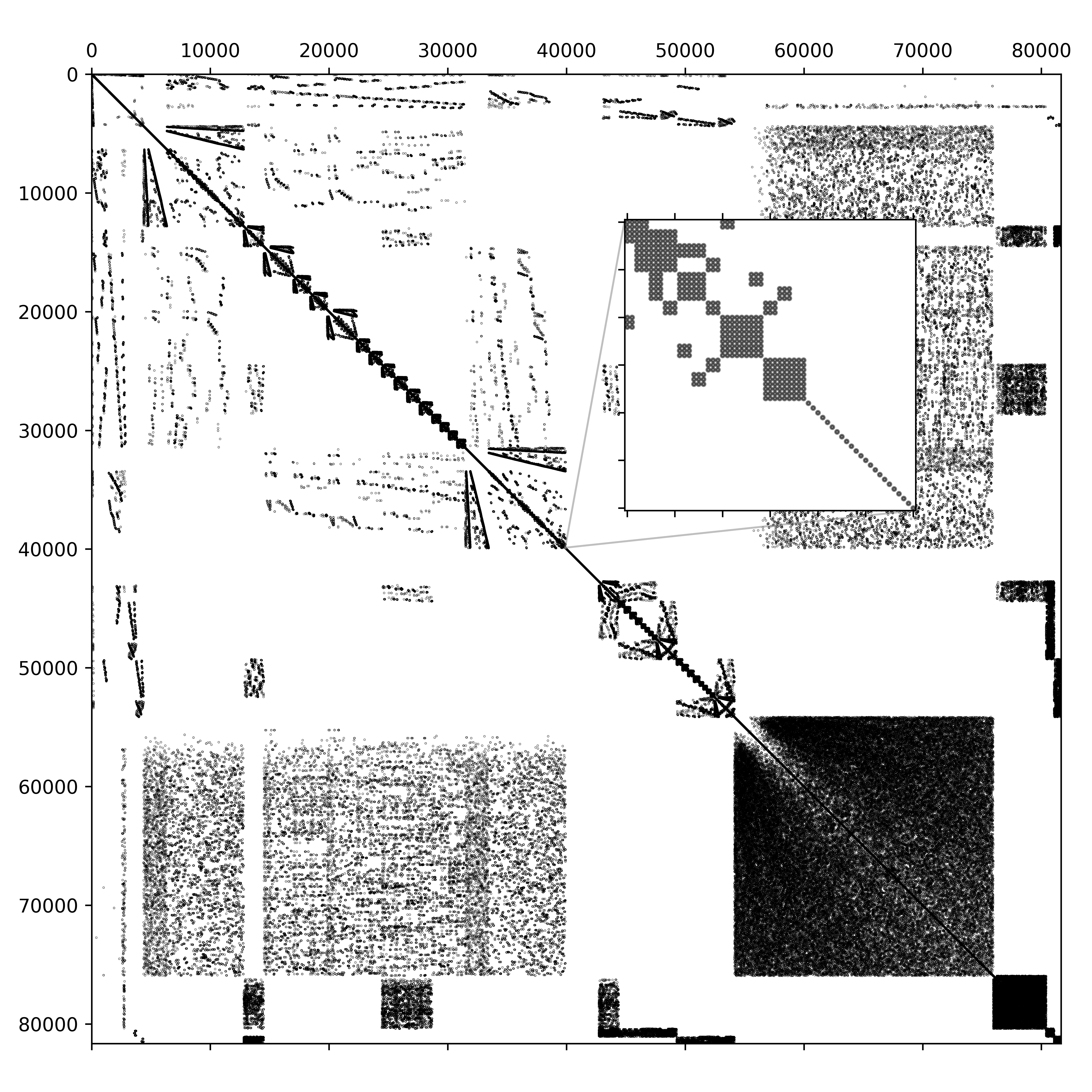}%
            \label{fig:A1}
        }
        \quad
        \subfigure[The second level of the AMG hierarchy (with the zero-energy modes provided).]{%
            \includegraphics[width=0.48\linewidth]{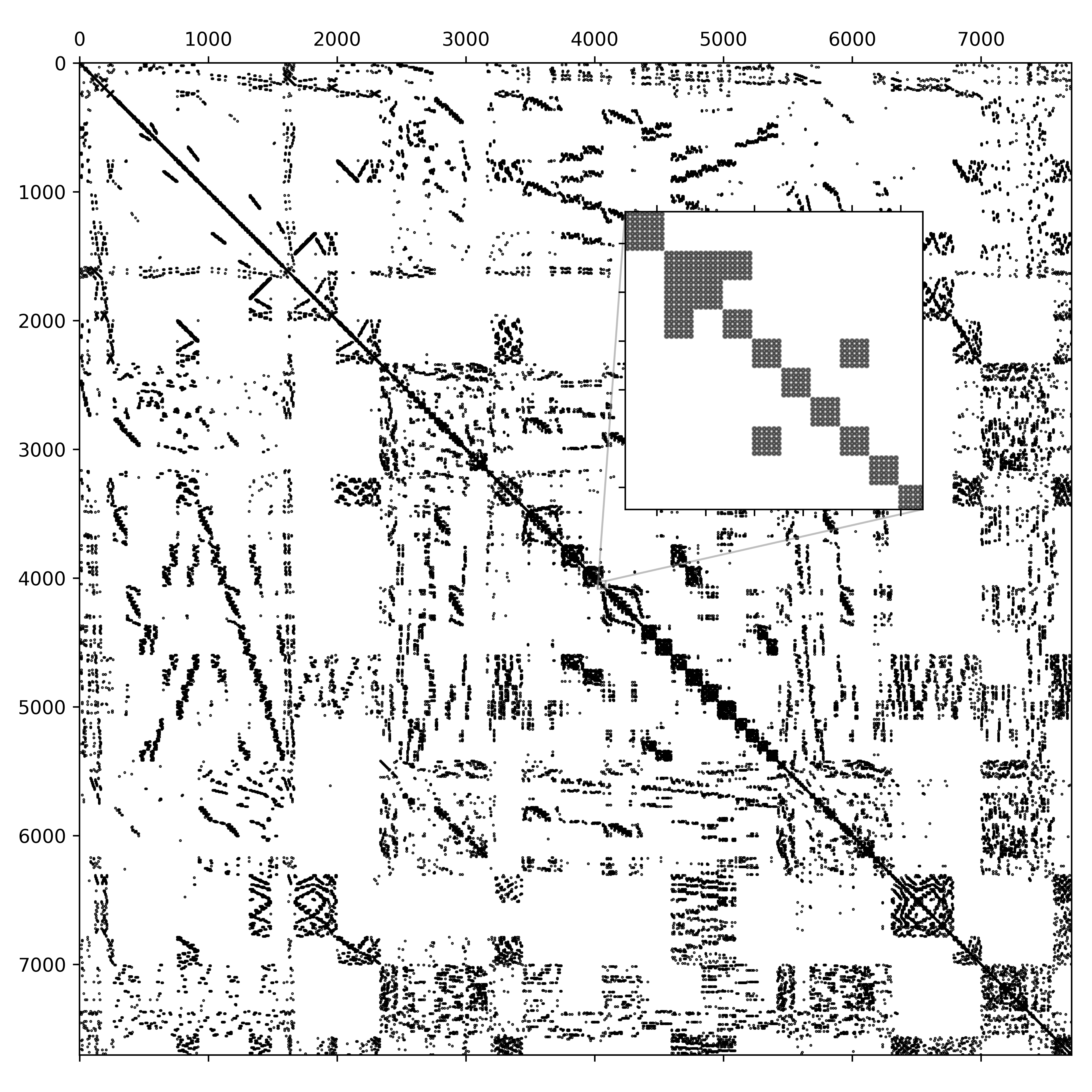}%
            \label{fig:A2}
        }
    \end{center}%
    \caption{Non-zero pattern of the system matrices for the connected rod
    problem from \Cref{sec:experiments}.}
    \label{fig:nzpattern}
\end{figure}

When the system matrix $A$ is obtained by discretization of a 3D elasticity
problem, it has block structure with small $3 \times 3$ blocks, as illustrated
by~\Cref{fig:A1} showing the non-zero pattern of the system matrix for the
connecting rod model. When the six rigid body modes are used as near nullspace
components, the matrices at the subsequent levels of the AMG hierarchy also
have block structure with $6 \times 6$ blocks, as shown on~\Cref{fig:A2}.
This means that the system matrices at every level of the constructed hierarchy
may be represented using the $3 \times 3$ block values.  However, even though
most of the steps in \Cref{alg:setup} and all of the steps in~\Cref{alg:vcycle}
may be expressed in terms of block arithmetics, the construction of the
tentative prolongation operator $P_i$ together with the next level matrix
$B_{i+1}$ has to be performed using scalar arithmetics. The next section
outlines the proposed workarounds for this issue.

\section{Enabling the use of block arithmetics in AMGCL}
\label{sec:nsblock}

The multigrid hierarchy in the AMGCL library is constructed using builtin data
structures and then transferred into one of the provided backends. This allows
for transparent acceleration of the solution phase with help of OpenMP, OpenCL,
or CUDA technologies. This may also be used for transparently conversion of the
constructed matrices at every level of the hierarchy into the block format. The
implementation is shown in \Cref{lst:hybrid}. Here, the hybrid backend extends
the builtin one and overrides the static function \code{copy_matrix()} that
transfers a matrix into the backend format. This way, the AMG hierarchy is
constructed in the scalar space, and the matrices are converted into the block
representation afterwards. In \Cref{sec:experiments} this method is shown as
\emph{NS Hybrid1}.

\begin{lstlisting}[float=t,
    caption={Implementation of the hybrid backend, converting the constructed
    matrices to the block format.},
    label=lst:hybrid]
template <class Block>
struct builtin_hybrid : public builtin<typename math::scalar_of<Block>::type>
{
    typedef builtin<typename math::scalar_of<Block>::type> Base;
    typedef crs<Block> matrix;
  
    static auto copy_matrix(
        std::shared_ptr<typename Base::matrix> As,
        const typename Base::params&
        )
    {
        return std::make_shared<matrix>(adapter::block_matrix<Block>(*As));
    }
};
\end{lstlisting}

This approach has a couple of disadvantages. First, the setup is still
performed using scalar arithmetics, and does not have any performance gains of
using block arithmetics. Second, it may work well with simpler relaxation
schemes, such as damped Jacobi or sparse approximate inverse, where the most
time consuming operation is a sparse matrix-vector product, which is performed
using block arithmetics automatically. But the more involved smoothers, such as
the ones based on incomplete LU factorization~\cite{saad2003iterative}, will
still be constructed and applied using scalar arithmetics, which means there
will be no performance gains for the most time-consuming part of the solution.

In order to deal with the latter drawback, a relaxation wrapper has to be
introduced. The wrapper will first convert the input matrix into the block
format and only then construct the underlying relaxation scheme. This is
implemented in AMGCL as \code{amgcl::relaxation::as_block} helper class
template. The implementation is straightforward and is not shown here for the
sake of brevity. The \emph{NS Hybrid2} solver in \Cref{sec:experiments} uses
the hybrid backend with the incomplete LU relaxation wrapped into the
\code{as_block} helper class.

\begin{lstlisting}[float=t,
    caption={Coarsening scheme wrapper that converts the input matrix from
    block into scalar format, applies the underlying coarsening scheme, and
    converts the results back to the block format.},
    label=lst:scalarcoarsening]
template <template <class> class Coarsening>
struct as_scalar {
    template <class Backend>
    struct type {
        typedef typename Backend::value_type Block;
        typedef typename math::scalar_of<Block>::type Scalar;
        typedef Coarsening< backend::builtin<Scalar> > Base;
  
        typedef typename Base::params params;
        Base base;
  
        type(const params &prm = params()) : base(prm) {};
        
        template <class Matrix>
        auto transfer_operators(const Matrix &B) {
            auto [P, R] = base.transfer_operators(*adapter::unblock_matrix(B));
  
            return std::make_tuple(
                    std::make_shared<Matrix>(adapter::block_matrix<Block>(*P)),
                    std::make_shared<Matrix>(adapter::block_matrix<Block>(*R)));
        }       
    };
};
\end{lstlisting}

The hybrid backend method is a half-measure in the sense that the
preconditioner setup is still performed using the scalar arithmetics. As was
discussed earlier, it should be possible to perform most of the setup
operations in the block space except for the computation of the transfer
operators. This may be implemented with the help of a coarsening scheme wrapper
\code{amgcl::coarsening::as_scalar} presented on \Cref{lst:scalarcoarsening}.
The wrapper converts the input system matrix from the block into the scalar
format, applies the underlying coarsening scheme in the scalar space, and
converts the computed transfer operators back into the block format. This way
only the coarsening step is performed in the scalar space, and all of the other
setup operations, including the time consuming Galerkin operator and the
relaxation scheme construction, are done using block arithmetics. This solver
is designated as \emph{NS Block} in the next section.

\section{Model problems and performance results} \label{sec:experiments}

\begin{figure}
    \begin{center}
        \subfigure[Connecting rod.]{%
            \includegraphics[width=0.45\linewidth]{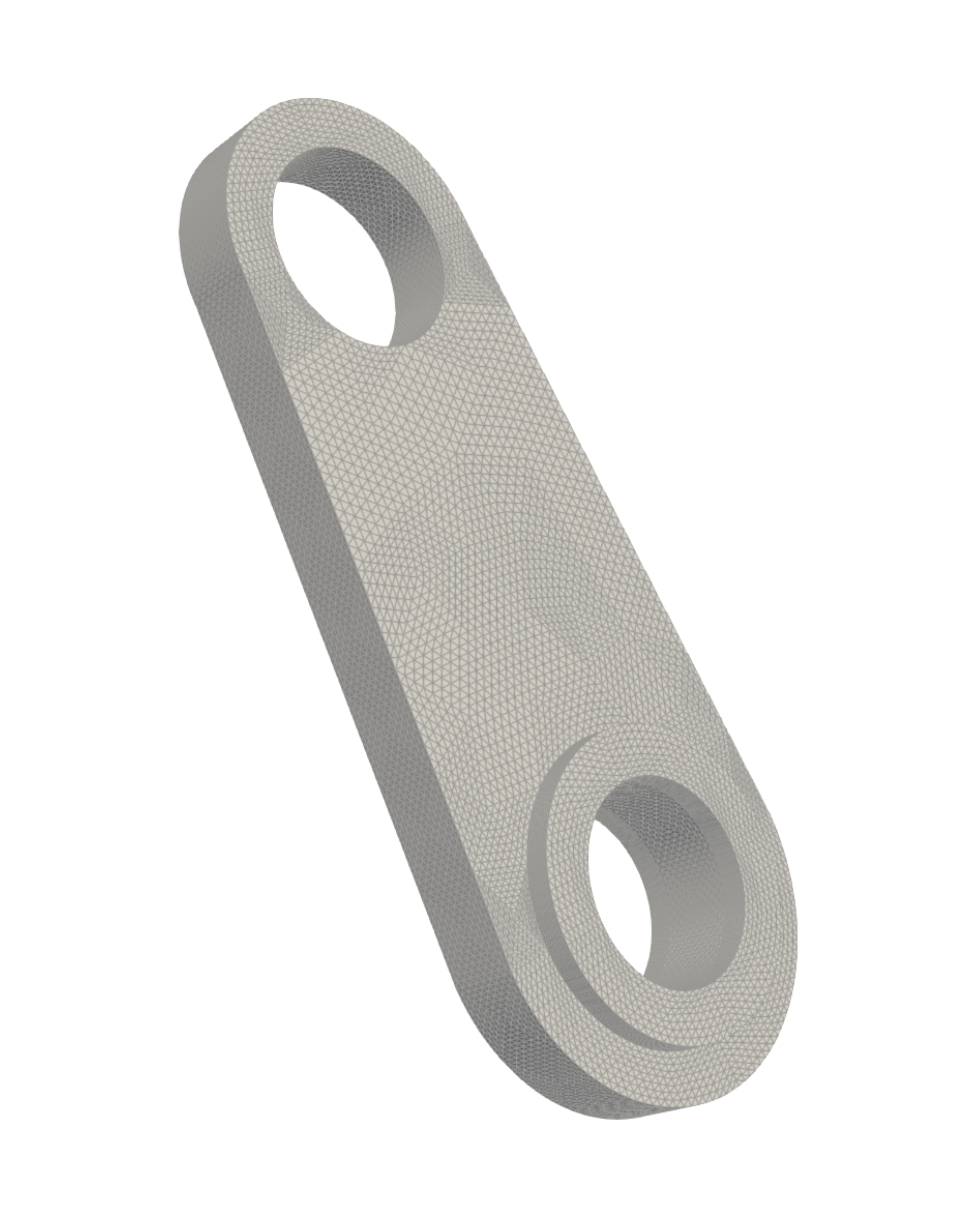}%
            \label{fig:connrod}
        }
        \subfigure[Solid hook.]{%
            \includegraphics[width=0.45\linewidth]{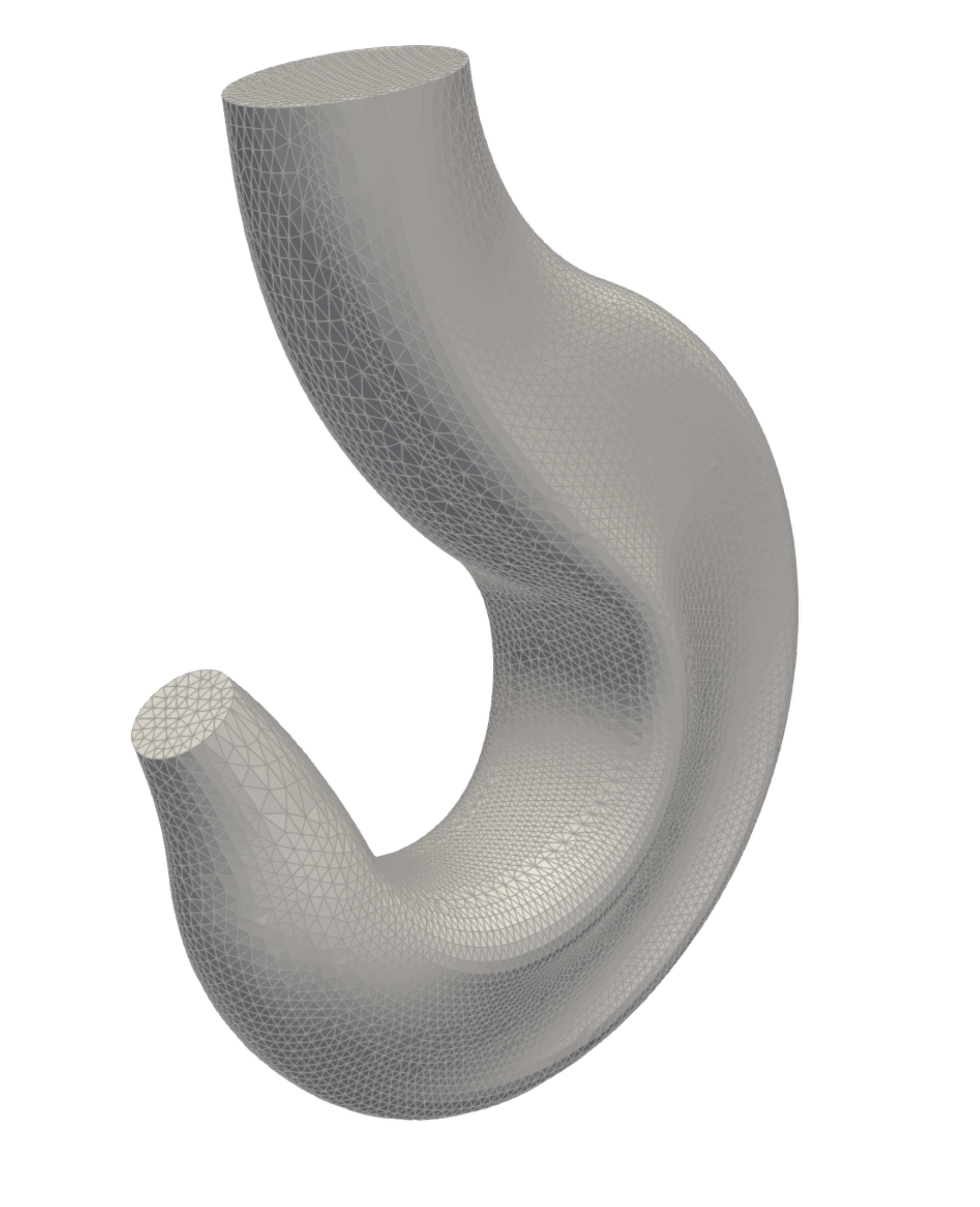}%
            \label{fig:hook}
        }
    \end{center}
    \caption{Geometries of the model problems.}
    \label{fig:models}
\end{figure}

The proposed techniques for enabling the use of block arithmetics with smoothed
aggregation AMG are benchmarked in this section on the examples of two 3D
elasticity problems. The first problem models a connecting rod shown
on~\Cref{fig:connrod}. The discretized system matrix has 81\,657 unknowns and
3\,171\,111 nonzero elements and was kindly provided by David Herrero-P\'erez
in~\cite{david_herrero_perez_2020_4299865}. The second problem is a solid hook
shape optimization example~\cite{geiser2021aggregated,kratos_3d_hook} from
Kratos Multi-Physics framework~\cite{Dadvand2010,Dadvand2013}. The system
matrix here is taken from the first iteration of the optimization algorithm and
contains 59\,718 unknowns and 2\,309\,832 nonzero elements. The solid hook
geometry is shown on~\Cref{fig:hook}.

Each of the model systems is solved with the six solvers outlined below. Every
version uses the conjugate gradient iterative solver with different variations
of the smoothed aggregation AMG preconditioner. The incomplete LU factorization
with zero fill-in (ILU(0)) is used as a smoother in every case. The relative
residual threshold of $10^{-8}$ is used in all experiments.  All tests were
conducted on a 3.40GHz Intel Core i5-3570K CPU. The complete source code for
the benchmarks is published in~\cite{bmcode}.

\emph{Scalar} solver. Here the preconditioner is a standard smoothed
aggregation AMG with the scalar double precision value type. Point-wise
aggregation with block size of 3 is used to account for the fact that the
system matrix has block structure. The near nullspace components for the
problem are not used. The definition of the solver is presented
in~\Cref{lst:scalar}. The \code{make_solver} class here binds together an
iterative solver (CG) and a preconditioner (AMG). Both the solver and the
preconditioner are parameterized on the backend, which in this case is the
builtin backend with plain \code{double} as value type. The AMG preconditioner
class has two other template parameters: coarsening (here
\code{smoothed_aggregation}) and relaxation (here \code{ilu0}).

\begin{lstlisting}[float=t,
    caption={Declaration of the Scalar solver.},
    label=lst:scalar
    ]
typedef amgcl::backend::builtin<double> Backend;
typedef amgcl::make_solver<
    amgcl::amg<
        Backend,
        amgcl::coarsening::smoothed_aggregation,
        amgcl::relaxation::ilu0
        >,
    amgcl::solver::cg<Backend>
    > Solver;
\end{lstlisting}

\emph{Block} solver. This is similar to the Scalar solver, except that the
block-valued backend is used instead of the scalar-valued one. Plain
aggregation is used as opposed to the point-wise one, since the coarsening with
block values automatically accounts for the block structure of the system
matrix. The solver is defined as shown in~\Cref{lst:block}.  Both the Scalar
and the Block solvers do not employ the near nullspace components of the
problem and are only used here to show that providing rigid body modes is
important to get good convergence in elasticity problems.

\begin{lstlisting}[float=t,
    caption={Declaration of the Block solver.},
    label=lst:block
    ]
typedef amgcl::static_matrix<double, 3, 3> Block;
typedef amgcl::backend::builtin<Block> Backend;
typedef amgcl::make_solver<
    amgcl::amg<
        Backend,
        amgcl::coarsening::smoothed_aggregation,
        amgcl::relaxation::ilu0
        >,
    amgcl::solver::cg<Backend>
    > Solver;
\end{lstlisting}

\emph{NS Scalar} solver uses scalar arithmetics throughout and is used as a
baseline reference for the rest of the solvers. The solver definition is
identical to the Scalar solver, but the rigid body modes are computed from the
grid coordinates and provided as near nullspace components to the solver
constructor. All the solvers below also use the near nullspace components to
construct the tentative prolongation operator.

\emph{NS Hybrid1} solver uses the hybrid backend introduced in the previous
section. Here, the preconditioner is set up using scalar arithmetics, and then
the constructed matrices are converted to the block format. The definition of
the solver is shown in~\Cref{lst:nshyb1}.

\begin{lstlisting}[float=t,
    caption={Declaration of the NS Hybrid1 solver.},
    label=lst:nshyb1
    ]
typedef amgcl::static_matrix<double, 3, 3> Block;
typedef amgcl::backend::builtin_hybrid<Block> Backend;
typedef amgcl::make_solver<
    amgcl::amg<
        Backend,
        amgcl::coarsening::smoothed_aggregation,
        amgcl::relaxation::ilu0
        >,
    amgcl::solver::cg<Backend>
    > Solver;
\end{lstlisting}

\emph{NS Hybrid2} solver differs from \emph{NS Hybrid1} in the choice of the
relaxation scheme. Here, the ILU(0) relaxation is wrapped into the
\code{relaxation::as_block} helper class, so that the relaxation is constructed
using block arithmetics. The solver is defined according to~\Cref{lst:nshyb2}.

\begin{lstlisting}[float=t,
    caption={Declaration of the NS Hybrid2 solver.},
    label=lst:nshyb2
    ]
typedef amgcl::static_matrix<double, 3, 3> Block;
typedef amgcl::backend::builtin_hybrid<Block> Backend;
typedef amgcl::make_solver<
    amgcl::amg<
        Backend,
        amgcl::coarsening::smoothed_aggregation,
        amgcl::relaxation::as_block<
            amgcl::backend::builtin<Block>,
            amgcl::relaxation::ilu0
            >::type
        >,
    amgcl::solver::cg<Backend>
    > Solver;
\end{lstlisting}

\emph{NS Block} solver uses the builtin block-valued backend similar to the
Block solver, but the coarsening scheme here is wrapped into the
\code{coarsening::as_scalar} helper class. This allows the complete
preconditioner setup to be performed in block space except for the coarsening,
which is done using scalar arithmetics. \Cref{lst:nsblock} shows the definition
of the solver in this case.

\begin{lstlisting}[float=t,
    caption={Declaration of the NS Block solver.},
    label=lst:nsblock
    ]
typedef amgcl::static_matrix<double, 3, 3> Block;
typedef amgcl::backend::builtin<Block> Backend;
typedef amgcl::make_solver<
    amgcl::amg<
        Backend,
        amgcl::coarsening::as_scalar<
            amgcl::coarsening::smoothed_aggregation
            >,
        amgcl::relaxation::ilu0
        >,
    amgcl::solver::cg<Backend>
    > Solver;
\end{lstlisting}

The performance results for the above solvers are shown
in~\Cref{tab:connrod,tab:hook}. The tables show the preconditioner setup time
and the solution time in seconds, the sum of the two times as the total
compute time, the number of iterations required for the solver to converge, and
the memory footprint of each preconditioner version in megabytes.

\begin{table}
    \caption{Performance results for the connecting rod model.}%
    \label{tab:connrod}
    \begin{tabularx}{\textwidth}{lYYYYY}
        \toprule
        Solver     & Setup (s)   & Solve (s)   & Total (s)   & Iterations & Memory (M) \\
        \midrule
        Scalar     &      0.330  &      1.942  &      2.271  &      96  &     138.86  \\
        Block      & {\bf 0.117} &      1.065  &      1.182  &      74  & {\bf 83.68} \\
        NS Scalar  &      0.673  &      0.771  &      1.444  & {\bf 29} &     205.16  \\
        NS Hybrid1 &      0.699  &      0.621  &      1.321  & {\bf 29} &     147.38  \\
        NS Hybrid2 &      0.454  & {\bf 0.501} &      0.955  & {\bf 29} &     114.19  \\
        NS Block   &      0.300  & {\bf 0.503} & {\bf 0.803} & {\bf 29} &     114.19  \\
        \bottomrule
    \end{tabularx}
\end{table}

\begin{table}
    \caption{Performance results for the solid hook model.}%
    \label{tab:hook}
    \begin{tabularx}{\textwidth}{lYYYYY}
        \toprule
        Solver     & Setup (s)   & Solve (s)   & Total (s)   & Iterations & Memory (M) \\
        \midrule
        Scalar     &      0.227  &      1.366  &      1.594  &      82  &     110.68  \\
        Block      & {\bf 0.067} &      0.672  &      0.739  &      73  & {\bf 61.89} \\
        NS Scalar  &      0.591  &      0.500  &      1.091  & {\bf 21} &     179.47  \\
        NS Hybrid1 &      0.613  &      0.421  &      1.034  & {\bf 21} &     129.52  \\
        NS Hybrid2 &      0.355  & {\bf 0.283} &      0.638  & {\bf 21} &      99.76  \\
        NS Block   &      0.230  & {\bf 0.279} & {\bf 0.509} & {\bf 21} &      99.76  \\
        \bottomrule
    \end{tabularx}
\end{table}


The Scalar solver, as expected, has the worst overall performance, since it
does not use any problem-specific information aside from the aggregation block
size. The Block solver does not rely on the near nullspace components
either and behaves similar to the Scalar solver in terms of convergence (the
number of iterations for the solver is only slightly lower than that of the
Scalar one).  However, using the block arithmetics allows to reduce the setup
time by a factor of about 3, and the total compute time is almost twice lower.
The memory footprint of the preconditioner is also significantly reduced. The
Block solver has the fastest setup time and the lowest memory footprint among
all of the considered solvers.

The NS Scalar solver uses the rigid body modes provided with each of the
models, but does not use the block arithmetics. The fact that the six near
nullspace components are used for the construction of the tentative
prolongation makes the AMG hierarchy much heavier in terms of the memory
footprint: the preconditioner requires 50--60\% more memory, and each iteration
of the solver takes about 30--40\% more time. However, the use of the rigid
body modes results in the significant improvement of the solver performance. NS
Scalar solver converges about 3 times faster and thus the overall compute time
is about 50\% lower than that of the Scalar solver.  Interestingly, even with
the improved convergence, the NS Scalar solver is still slower than the simple
Block solver. Due to the much faster setup the Block solver is able to
outperform the NS Scalar solver by 20--30\%. This makes the Block solver a
viable choice in cases where the problem is able to converge at all without the
use of the near nullspace components.

NS Hybrid1 solver constructs the preconditioner in the scalar space and
converts the constructed matrices into the block representation. Hence, its
setup time is slightly higher than that of the NS Scalar solver. The solution
time for the NS Hybrid1 is lower than NS Scalar, but not in a significant way.
This is explained by the fact that about 75\% of the solution time is spent on
relaxation, and the ILU(0) relaxation here is both constructed and applied in
the scalar space. This issue is resolved in the NS Hybrid2 solver, which uses
\code{as_block} relaxation wrapper to move the construction and application of
the ILU(0) relaxation into the block space. This reduces the setup time by
35--40\%, and improves the solution time by 20--30\% compared to the NS Hybrid1
solver, and by 35--40\% compared to the NS Scalar solver. The memory footprint
of the NS Hybrid2 solver is also reduced by 20--40\%.

The best overall performance is shown by the NS Block solver. Here, the
preconditioner is almost completely constructed using block arithmetics with
the exception of the transfer operators computation, which is done in the
scalar space with the help of the \code{coarsening::as_scalar} wrapper class.
The setup time here is the best among the NS family of solvers, and is on par
with the simple Scalar solver. The solution time is similar to the NS Hybrid2
solver, since both solvers, even though slightly different in terms of
construction, result in identical internal data structures. This is confirmed
by the fact that both solvers have equivalent memory footprints. In terms of
the overall compute time the NS Block solver is able to outperform the NS
Scalar solver by almost 50\%, and the simple Scalar solver~--- by about 65\%.

\section{Summary} \label{sec:summary}

It has been shown that using block arithmetics may significantly speed up the
solution of 3D elasticity problems with smoothed aggregation AMG. Two practical
techniques of enabling the use of block arithmetics based on the open-source
AMGCL library have been suggested. The first approach introduces a hybrid
backend that constructs the complete AMG hierarchy using scalar values, and
then converts the computed matrices into the block representation. The second
technique uses block-valued backend with a coarsening wrapper that converts the
input matrix into the scalar format, applies the standard coarsening using
scalar arithmetics, and converts the computed transfer operators back to the
block representation. The second technique turned out to be more effective. The
NS Block solver is able to outperform the fully scalar method by about 50\% and
requires about 30\% less memory for the preconditioner.  Another advantage of
the NS Block solver is that it does not need to introduce another backend, and
the same code may be used with OpenMP or GPGPU backends. Both techniques have
been made available to the scientific community as part of the open-source
AMGCL library. The flexibility of the library allowed to implement the
suggested methods with a minimal amount of code.

\section{Acknowledgements}

The work was carried out at the JSCC RAS as part of the government assignment.


\begin{thebibliography}{10}
\providecommand{\url}[1]{{#1}}
\providecommand{\urlprefix}{URL: }
\expandafter\ifx\csname urlstyle\endcsname\relax
  \providecommand{\doi}[1]{DOI:~\discretionary{}{}{}#1}\else
  \providecommand{\doi}{DOI:~\discretionary{}{}{}\begingroup
  \urlstyle{rm}\Url}\fi

\bibitem{barrett1994templates}
R.~Barrett, M.~Berry, T.F. Chan, J.~Demmel, J.~Donato, J.~Dongarra,
V.~Eijkhout, R.~Pozo, C.~Romine, H.~Van~der Vorst, {\em Templates for the solution
of linear systems: building blocks for iterative methods.}
\newblock SIAM,  1994.

\bibitem{brandt1985algebraic}
A.~Brandt, S.~McCormick, J.~Huge, {\em Algebraic multigrid ({AMG}) for sparse matrix
equations.}
\newblock {\em Sparsity and its Applications} \textbf{257} (1985).

\bibitem{Dadvand2013}
P.~Dadvand, R.~Rossi, M.~Gil, X.~Martorell, J.~Cotela, E.~Juanpere, S.R.
  Idelsohn, E.~O{\~{n}}ate, {Migration of a generic multi-physics framework to
  HPC environments}.
\newblock Computers and Fluids \textbf{80}(1), 301--309 (2013).
\newblock \doi{10.1016/j.compfluid.2012.02.004}.

\bibitem{Dadvand2010}
P.~Dadvand, R.~Rossi, E.~O{\~{n}}ate, {An object-oriented environment for
  developing finite element codes for multi-disciplinary applications}.
\newblock Archives of Computational Methods in Engineering \textbf{17}(3),
  253--297 (2010).
\newblock \doi{10.1007/s11831-010-9045-2}.

\bibitem{Demidov2019}
D.~Demidov, {AMGCL}: An efficient, flexible, and extensible algebraic multigrid
  implementation.
\newblock Lobachevskii Journal of Mathematics \textbf{40}(5), 535--546 (2019).
\newblock \doi{10.1134/S1995080219050056}.

\bibitem{Demidov2020}
D.~Demidov, {AMGCL} -- a {C++} library for efficient solution of large sparse
  linear systems.
\newblock Software Impacts \textbf{6}, 100037 (2020).
\newblock \doi{10.1016/j.simpa.2020.100037}.

\bibitem{bmcode}
D.~Demidov, {Source code for the benchmarks in "Efficient solution of 3D
  elasticity problems with smoothed aggregation algebraic multigrid and block
  arithmetics"}.
\newblock \url{https://github.com/ddemidov/block_nullspace_benchmarks} (2022).
\newblock Accessed 17.02.2022.

\bibitem{DEMIDOV2021101285}
D.~Demidov, L.~Mu, B.~Wang, Accelerating linear solvers for {Stokes} problems
  with {C++} metaprogramming.
\newblock Journal of Computational Science \textbf{49}, 101285 (2021).
\newblock \doi{https://doi.org/10.1016/j.jocs.2020.101285}.
\newblock
  \urlprefix\url{https://www.sciencedirect.com/science/article/pii/S1877750320305809}.

\bibitem{kratos_3d_hook}
A.~Geiser, Optimization of a solid {3D} hook subjected to multiple constraints.
\newblock
  \urlprefix\url{https://github.com/KratosMultiphysics/Examples/tree/master/shape_optimization/use_cases/10_Multi_Constraint_Optimization_3D_Hook}.
\newblock Accessed 08.02.2022.

\bibitem{geiser2021aggregated}
A.~Geiser, I.~Antonau, K.U. Bletzinger, Aggregated formulation of geometric
  constraints for node-based shape optimization with vertex morphing.
\newblock In: 14th International Conference on Evolutionary and Deterministic
  Methods for Design, Optimization and Control, pp. 80--94 (2021).
\newblock \doi{10.7712/140121.7952.18383}.

\bibitem{gupta2010adaptive}
A.~Gupta, T.~George, Adaptive techniques for improving the performance of
  incomplete factorization preconditioning.
\newblock SIAM Journal on Scientific Computing \textbf{32}(1), 84--110 (2010).
\newblock \doi{10.1137/080727695}.

\bibitem{david_herrero_perez_2020_4299865}
D.~Herrero-P{\'e}rez, Discretization of a {3D} elasticity problem (2020).
\newblock \doi{10.5281/zenodo.4299865}.
\newblock \urlprefix\url{https://doi.org/10.5281/zenodo.4299865}.

\bibitem{saad2003iterative}
Y.~Saad, {\em Iterative methods for sparse linear systems}.
\newblock Siam,  2003.

\bibitem{Stuben1999}
K.~Stuben, {Algebraic multigrid (AMG): an introduction with applications}.
\newblock GMD Report~70, GMD, Sankt Augustin, Germany (1999).

\bibitem{Trottenberg2001}
U.~Trottenberg, C.~Oosterlee, A.~Sch{\"{u}}ller, {\em Multigrid}.
\newblock Academic Press, London,  2001.

\bibitem{VanekBrezinaMandel2001}
P.~Van{\v{e}}k, M.~Brezina, J.~Mandel, Convergence of algebraic multigrid based
  on smoothed aggregation.
\newblock Numer.Math. \textbf{88}, 559--579 (2001).
\newblock \doi{10.1007/s211-001-8015-y}.

\end{thebibliography}

\end{document}